\documentclass[aps,twocolumn,prd,showpacs,nofootinbib]{revtex4}
\usepackage{amsmath}
\usepackage{graphicx}
\usepackage{dcolumn}
\usepackage{bm}
\usepackage{amssymb}
\usepackage{latexsym}

\def\be{\begin{equation}}
\def\ee{\end{equation}}
\def\ba{\begin{eqnarray}}
\def\ea{\end{eqnarray}}

\begin{document}

\title{A Galileon Design of Slow Expansion: II }

\author{Zhi-Guo Liu }
\author{Yun-Song Piao}

\affiliation{College of Physical Sciences, Graduate University of
Chinese Academy of Sciences, Beijing 100049, China}

\begin{abstract}

We parameterize the evolutions of the slow expansion, which
emerges from a static state in infinite past, into different
classes. We show that the scale invariant adiabatical perturbation
may be generated during these evolutions, and the corresponding
evolutions can be realized with generalized Galileon Lagrangians,
in which there is not the ghost instability.

\end{abstract}

\maketitle

\section{Introduction}

How to apprehend the beginning of the hot ``big bang" model has
been still a significant issue. The inflationary scenario
\cite{G},\cite{LAS},\cite{S80} is the current paradigm of the
early universe, which not only homogenizes the universe but
provides the scale invariant primordial perturbation responsible
for the observable universe \cite{MC}. However, the universe still
requires a beginning, since in inflationary scenario any backward
null or timelike geodesic has a finite affine length
\cite{Borde:2001nh}.

The idea of the emergent universe is interesting
\cite{Ellis:2002we},\cite{Ellis:2003qz}, in this scenario the
universe originates from a static state in the infinite past.
In emergent universe scenario, it seems that there might be not a
beginning, since the affine length of backward null or timelike
geodesic is classical infinite. However, for the model, in which
the initial static state is constructed by applying a positive
curvature, the static universe will inevitably collapse quantum
mechanically \cite{Vilenkin11},\cite{Dabrowski}.

In Refs.\cite{Ellis:2002we},\cite{Ellis:2003qz}, after the
universe emerges from the static state, the inflation will be
required to set the initial conditions of ``big bang" model,
i.e.the homogenizing and the scale invariant primordial
perturbation.

However, when the universe emerges, or begins to deviate from
static state, it might be slowly expanding. In Ref.\cite{PZhou},
it has been for the first time observed that the slow expansion
might adiabatically generate the scale invariant curvature
perturbation, see \cite{Piao0706} for that induced by the entropy
perturbation. Thus it might be imaginable that in emergent
universe scenario, the initial state of ``big bang" evolution
could be set during this slowly expanding period. During the slow
expansion, $\epsilon<0$ is required \cite{PZhou},\cite{Piao1012}.
Thus in Ref.\cite{PZhou}, the phantom has been applied for a
phenomenological studying.







Recently, the cosmological application of Galileon, \cite{NRT}, or
its nontrivial generalization \cite{Vikman},\cite{KYY},\cite{DGS},
has acquired increasing attentions, e.g. see
\cite{Vikman},\cite{G1} for dark energy, \cite{KYY},\cite{Seery}
for inflation, \cite{Gcurvaton} for curvaton and \cite{Gbounce}
for bouncing universe, in which $\epsilon<0$ can be implemented
stably, there is not the ghost instability.

The models of the emergent universe scenario can be realized with
Galileon, e.g.\cite{CNT}, and also \cite{Liu:2011ns}. However, in
Refs.\cite{CNT},\cite{Wangyi}, the adiabatic perturbation is not
scale invariant, thus the obtaining of the scale invariant
curvature perturbation has to appeal to the conversion of the
perturbations of other light scalar fields, i.e. the mechanism
similar to that in Refs.\cite{Rubakov},\cite{HK}. However, we
showed that actually the scale invariant curvature perturbation
may be adiabatically generated \cite{Liu:2011ns}. Here, with
Ref.\cite{Borde:2001nh}, we will make the relevant scenario
clearer.

Here, we will parameterize the evolutions of the slow expansion,
which emerges from a static state in infinite past, into different
classes,
and clarify how the adiabatical perturbation generated could be
scale invariant for these parameterizations. When the slow
expansion phase ends, the universe reheats and the evolution of
``big bang" model begins. We have showed in Ref.\cite{Liu:2011ns}
that one of these parameterizations can be realized by applying a
generalized Galileon. Here, we will show that the other
parameterization can be also realized similarly, and the results
may be consistent with the observations.




\section{As A General Result}

\subsection{The parameterization of ``emergence" or slow expansion}

That in infinite past the universe is a static state, in which
$l_{init}>l_P$ is constant, $l_P=1/M_P$ and $l_{init}$ is the
physical size of initial universe, requires \be H\longrightarrow
0,\,\,\, when \,\,\, t\longrightarrow -infinity.
\label{Hrequire}\ee Thus the corresponding evolutions can be
parameterized as \ba H & \sim & { 1\over
\beta^b(t_*-t)^{b+1}},\,\,\, \label{H1}\\ & or &
e^{-\beta(t_*-t)}, \,\,\, \label{H2}\ea where $t<t_*$, and $b>0$
and $\beta\sim 1/|t_*|>0$ are constant. Thus in infinite past the
behavior of $H\longrightarrow 0$ is power law or is exponential.
However, of course, the behavior of $H$ could be
also double exponential \ba H & \sim & e^{-e^{\beta(t_*-t)}}, \label{H3}\\
& or & \,\,{\rm higher}\,\, {\rm exponential}.\ea Here, we will be
only limited to (\ref{H1}) and (\ref{H2}). Thus $a=e^{\int Hdt}$
is given by \ba a
& \sim & e^{ 1/[\beta^b(t_*-t)^{b}]},\,\,\, ({\rm parameterization}\,\,{\rm I}) \label{model1}\\
& or & e^{e^{-\beta(t_*-t)}}, \,\,\, ({\rm
parameterization}\,\,{\rm II}), \label{model2}\ea where
$a_{init}=1$ is set in the infinite past. When $t_f\simeq {\cal
O}(1)t_*$, $a \simeq e$ for (\ref{model1}) and $a\simeq e^{1/e}$
for (\ref{model2}). Thus in the regime $-infinite< t<t_f$, the
universe is slowly expanding. Thus we may define the ``emergence"
as a period of slow expansion, which emerges from a static state
in infinite past.

In Ref.\cite{Borde:2001nh}, it has been showed that in a
cosmological scenario, if $H_{average}
> 0$, this scenario will be incomplete in the infinite past, since any backward null geodesic will
have a finite affine length, i.e. a beginning is required. Here,
$H_{average}$ is an average over the affine parameter, and is
defined as \cite{Borde:2001nh} \be
H_{average}={\int^{t_f}H(\lambda)d\lambda \over
\int^{t_f}d\lambda} < {1\over \int^{t_f}d\lambda}, \label{HA}\ee
where $\lambda$ is the affine parameter of the backward null
geodesic, $d\lambda = adt/{a(t_f)}$, which gives \be
\int^{t_f}H(\lambda)d\lambda =\int^{a(t_f)}da <1. \label{Hda}\ee
Eq.(\ref{HA}) is universal, independent of models. Thus with
(\ref{Hda}), $H_{average}>0$ implies that the affine length $\int
d\lambda$ must be finite.

We can calculate the affine length of the backward null geodesic
for the slow expansion, which is parameterized as (\ref{model1})
or (\ref{model2}), which is \ba \int^{t_f} d\lambda  & = &
\int^{t_f} {a\over {a(t_f)}} dt,\nonumber\\ &\sim & {e^{x_1}\over
x_1^{1+1/b}}\mid_0^{x_{1f}}+\int_0^{x_{1f}} {1\over
x_{1}^{2+1/b}}e^{x_1}dx_1, \label{lambda2}\\ & or & {e^{x_2}\over
x_2}\mid_0^{x_{2f}}+\int_0^{x_{2f}} {1\over x_{2}^{2}}e^{x_2}dx_2,
\label{lambda1}\ea where \be x_1={1\over \beta^b(t_*-t)^b},\,\,\,
x_2=e^{-\beta (t_*-t)}.\ee We see that both affine lengths are
diverged, which indicates that the emergent universe scenario,
parameterized as (\ref{model1}) or (\ref{model2}), could be
complete in the infinite past, and there is not the beginning.
When $b\gg 1$, the result of (\ref{lambda1}) is the same with that
of (\ref{lambda2}), thus the affine length of the backward null
geodesic in parameterization (\ref{model1}) is same with that in
parameterization (\ref{model2}), which implies that (\ref{model2})
is actually the limit case of (\ref{model1}).

Here, the definition of $\epsilon$ is ${d\over dt}({1/ H})$. Thus
in the slowly expanding phase $\epsilon<0$, since $H$ is rapidly
increasing. In infinite past, \ba |\epsilon |
 & \sim & \beta^b (t_*-t)^b,\,\,\, for\,\,{\rm parameterization}\,\,{\rm I}, \label{e1}\\
& or &  e^{\beta(t_*-t)},\,\,\, for\,\,{\rm
parameterization}\,\,{\rm II} \label{e2} \ea is diverged. This is
actually a result of the condition (\ref{Hrequire}), i.e. in
infinite past the universe is a static state, in which
$l_{init}>l_P$ is constant.

What if $\epsilon$ is not divergent.
When $|\epsilon|>0$ tends to be constant in infinite past, we have
\be H\sim {1\over |\epsilon| (t_*-t)}\longrightarrow 0,
\,\,\,{in}\,\,{\rm infinite}\,\,{\rm past},\ee which seems satisfy
(\ref{Hrequire}). However, \be a\sim e^{\int Hdt}\sim {1\over
(t_*-t)^{1/|\epsilon|}}\longrightarrow 0 \ee in the meantime,
which is not consistent with the requirement of $l_{init}>l_P$. In
principle, $l_{init}$ should be larger than $l_P$, thus
$a\longrightarrow 0$ in infinite past in certain sense implies
that the corresponding evolution requires a beginning.

When $|\epsilon|\longrightarrow 0$ in infinite past,
we have $b<0$ for the parameterization (\ref{model1}), thus \ba a
\sim e^{ -|b|\beta^{|b|}(t_*-t)^{|b|}} \longrightarrow
0,\,\,\,{in}\,\,{\rm infinite}\,\,{\rm past}, \ea which is also
not consistent with the requirement of $l_{init}>l_P$, as has been
argued.


The parameterization (\ref{model1}) has been implemented by
applying Galileon Lagrangians. In
Refs.\cite{Piao1012},\cite{Liu:2011ns}, it has been showed that
the adiabatical perturbation is scale invariant requires $b=4$,
i.e. $a$ behaviors as \be a\sim e^{1\over (t_*-t)^4}. \ee In
Ref.\cite{CNT}, $ a\sim e^{1\over (t_*-t)^2}$, thus the adiabatic
perturbation is not scale invariant.


While for the parameterization (\ref{model2}), $e^{-\beta
(t_*-t)}\sim 0$ in infinite past i.e. $\beta (t_*-t)\gg 1$, thus
we have \be a \sim e^{e^{-\beta(t_*-t)}}\sim 1+e^{\beta(t-t_*)},
\ee which is just that in original emergent universe scenario
\cite{Ellis:2002we},\cite{Ellis:2003qz}, in which the initial
static state is constructed by introducing a positive curvature.
In Ref.\cite{Ellis:2002we}, initially $\beta (t_*-t)\gg 1$, the
universe is slowly expanding from the static state, after the
slowly expanding phase ends, a period of inflation is required,
and the scale invariant adiabatical perturbation is generated
during inflation. Here, we will show that the scale invariant
adiabatical perturbation may be actually generated during the slow
expansion without the help of inflation and the corresponding
evolutions can be realized with generalized Galileon Lagrangians.


\subsection{How the adiabatic perturbation is scale invariant ?}

We will clarify how the adiabatical perturbation generated during
the slow expansion, parameterized as (\ref{model1}) or
(\ref{model2}), could be scale invariant. The quadratic action of
the curvature perturbation $\cal R$ is
\cite{GM},\cite{Vikman},\cite{KYY}, \be S_2\sim \int d\eta d^3x
{a^2 M_P^2 Q\over c_s^2}\left({{\cal R}^\prime}^2-{c_s^2}(\partial
{\cal R})^2\right), \ee where $Q>0$ and $c_s^2>0$ should be
satisfied for the avoidance of the ghost and gradient
instabilities.

$Q= M_P^2\epsilon$ for $P(X,\phi)$ \cite{GM}, but $Q$ is
complicated for Galileon \cite{Vikman},\cite{KYY}. However, as has
been confirmed in Ref.\cite{Liu:2011ns} and will be confirmed
again here, $Q\sim |\epsilon|$ for the slow expansion with
$\epsilon<0$.

The equation of $\cal R$ is
\be \tilde{u}_k^{\prime\prime}
+\left(k^2-{\tilde{z}^{\prime\prime}\over \tilde{z}}\right)
\tilde{u}_k = 0, \label{uk}\ee after defining $\tilde{u}_k \equiv
\tilde{z}{\cal R}_k$, where $'$ is the derivative for $y=\int c_s
d\eta$, $\tilde{z}\equiv a\sqrt{2M_P^2 Q/ c_s}$. When $k^2\ll
\tilde{z}^{\prime\prime}/\tilde{z}$, the solution of $\cal R$
given by Eq.(\ref{uk}) is \ba {\cal R} & \sim & C\,\,\,\,\,
is\,\,\,{{\rm constant}}\,\,\,{ {\rm mode}}\label{C}\\ &or &\,
D\int {dy\over \tilde{z}^2}\,\,\,\,\, is\,\,\,{{\rm
changed}}\,\,\,{ {\rm mode}} , \label{D}\ea where $D$ mode is
increasing or decaying dependent of the evolution of $\tilde{z}$.

The scale invariance of $\cal R$ requires
${\tilde{z}^{\prime\prime}\over \tilde{z}}\sim {2\over
(y_*-y)^2}$, which implies \ba \tilde{z}\sim a\sqrt{Q\over c_s}
&\sim & {1\over y_*-y}\,\,\, {for}\,\,\, {{\rm constant}}\,\,\,{
{\rm mode}}
\label{z2}\\
&or & (y_*-y)^2 \,\,\, {for}\,\,\,{ {\rm increasing}}\,\,\,{{\rm
mode}} \label{z1}\ea has to be satisfied.
In principle, $a$, $Q$ and $c_s$ can be changed, all together
contribute the change of $\tilde{z}$. When $a$ is rapidly changed
while others are hardly changed, the inflationary background and
the contraction dominated by the matter \cite{Wands99},\cite{S1}
are obtained, respectively.

\begin{figure}[t]
\includegraphics[width=7cm]{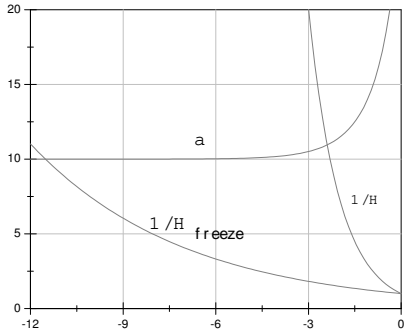}
\caption{The evolutions of $a$, the Hubble horizon and the $\cal
R$ horizon during the ``emergence" given by the parameterization
(\ref{model2}). $a_{init}=10$ is set. During this phase, due to
the rapid decreasing of $1/H$ and $1/H_{freeze}$, the adiabatical
perturbation mode initially deep inside both horizons, i.e.
$\lambda\sim a\ll 1/H_{freeze}\ll 1/H$, will leave the horizons,
and becomes the primordial perturbation responsible for the
observable universe.}
\end{figure}

When $Q$ is rapidly changed while $a$ is hardly changed, the scale
invariant adiabatical perturbation can be induced similarly by
either its constant mode \cite{KS1}, also criticism for it
\cite{LMV}, or its increasing mode \cite{Piao1012}.

We have showed that for the parameterization (\ref{model1}), the
scale invariance of adiabatical perturbation requires
\cite{Piao1012},\cite{Liu:2011ns}, \be a\sim e^{1\over (t_*-t)^4},
 \ee i.e.$|\epsilon|\sim (t_*-t)^4$, in which $c_s^2$ is constant.

Here, we will detailed clarify how to obtain scale invariant
adiabatical perturbation for the parameterization (\ref{model2}).
The scale invariance of the increasing mode of $\cal R$ requires
\be {Q}\sim  {c_s\over a^6}\left(\int c_s dt \right)^4,
\label{Qcs}\ee
where $dt\sim d\eta$, since $a$ is almost constant. Thus \be
c_s^2\sim e^{{2\over 5}\beta (t_*-t)}, \label{cs2}\ee where $Q\sim
|\epsilon|$ and (\ref{e2}) are applied. This implies that $c_s^2$
has to be infinite large in infinite past, and then rapidly
decreased.

In Ref.\cite{Picon}, it has been observed that the rapid running
of $c_s^2$ helps to obtain the scale invariant adiabatical
perturbation in noninflationary background. In
Refs.\cite{K1},\cite{JK2}, this issue is estimated again. In
Ref.\cite{K2}, it has been argued that the adiabatical
perturbation in the model of the slow expansion with rapidly
decreasing $c_s^2$ is not scale invariant. However, in their
settings, the adiabatical perturbation is dominated by its
constant mode, while here the perturbation is dominated by its
increasing mode.

When $k^2\simeq \tilde{z}^{\prime\prime}/\tilde{z}$, the
perturbation mode of $\cal R$ leaves its horizon, which is called
the $\cal R$ horizon \ba a/{\cal H}_{freeze} & = &
\sqrt{\left|{\tilde{z}\over
\tilde{z}^{\prime\prime}}\right|}\simeq \int c_s dt\nonumber\\
&\sim & {5\over \beta}e^{{1\over 5}\beta (t_*-t)}. \ea While the
Hubble horizon is $1/H$. Thus the evolutions of the $\cal R$
horizon and the Hubble horizon are different. We have \be a/ {\cal
H}_{freeze}\sim \left({1/ H}\right)^{1/5}. \label{HHH}\ee In
Ref.\cite{Liu:2011ns}, it has been showed how the scale invariant
adiabatical perturbation is obtained for (\ref{model1}), the
result also implies (\ref{HHH}) is satisfied. The evolutions of
both the $\cal R$ horizon and the Hubble horizon are same only
when $a$ is rapidly changed and $|\epsilon|$ is unchanged,
e.g.inflation, since $z^{\prime\prime}/z\sim a^{\prime\prime}/a$.

When $k^2\gg \tilde{z}^{\prime\prime}/\tilde{z}$, i.e. the
perturbation is deep inside the $\cal R$ horizon, $\tilde{u}_k$
oscillates with a constant amplitude. The quantization of
$\tilde{u}_k$ is well defined since $Q\sim |\epsilon|>0$, which
gives its initial value.
\be \tilde{u}_k\sim \frac{1}{\sqrt{2k}} \, e^{-ik\int c_s dt}. \ee
Thus initially \ba k^{3/2}{\cal R} & = &
k^{3/2}\left|{\tilde{u}_k\over \tilde{z}}\right|\sim k
\sqrt{c_s\over Q}\nonumber\\ &\sim & k e^{-{4\over
5}\beta(t_*-t)}\longrightarrow 0. \ea This insures that initially
the background of static state is not spoiled by the
perturbations. Thus in slow expansion scenario
\cite{PZhou},\cite{Liu:2011ns},\cite{LST}, the primordial
perturbation must be induced by the increasing mode, or it is
hardly possible to consistently define the initial perturbation in
the infinite past.


When $k^2\ll \tilde{z}^{\prime\prime}/\tilde{z}$, $\tilde{u}_k$ is
\ba \tilde{u}_{k}& = & {\sqrt{\pi}\over 2}e^{i\pi}\sqrt{-k\int c_s
dt} H_{3/2}^{(1)}(-k\int c_s dt)\nonumber\\ &\simeq & {1\over
\sqrt{2k}}\left(-k\int c_s dt\right)^{-1} . \ea Thus the amplitude
of perturbation spectrum is $ {\cal P}^{1/2}_{\cal R} \simeq
\sqrt{k^3}\left|{\tilde{u}_k\over \tilde{z}}\right|$. The
perturbation is given by the increasing mode (\ref{D}), which
implies that the spectrum of $\cal R$ should be calculated around
$t_{f}$.
Thus \be {\cal P}^{1/2}_{\cal R} \sim {1\over M_p \int c_s dt}
\sqrt{c_s\over Q}\sim {\beta\over M_p}. \label{P1}\ee The universe
will reheat around or after $t_{f}$, which has been studied in
\cite{Liu:2011ns}. Hereafter, the evolution of hot ``big bang"
model begins, the perturbation mode outside of horizon will
preserve constant.

Thus during the slow expansion, which emerge from a static state
in infinite past, the initial conditions of hot ``big bang"
evolution could be set.

\section{A Galileon Design of Slow Expansion:II}

In Ref.\cite{Liu:2011ns}, we have showed that the parameterization
(\ref{model1}) can be realized in the effective field theory
setting, in which there is not the ghost instability and the
perturbation spectrum is consistent with the observations. Here,
we will show the parameterization (\ref{model2}) can be also
realized similarly.

\subsection{The background}

\begin{figure}[t]
\includegraphics[width=7cm]{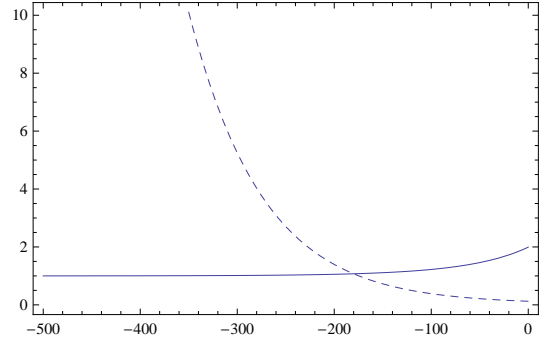}
\caption{We numerically solve Eqs.(\ref{phi}) and (\ref{H}), in
which the parameter ${\cal M}=0.01$. The solid line is the
evolution of $a$ and the dashed line is the evolution of $1/H$.
}\label{fig:aH}
\end{figure}

We consider a generalized Galileon as follows \be {\cal L}\sim
-\,{\varphi^5\over {\cal M}^5}\,X+{1\over {\cal
M}^{10}}X^{7/2}-{1\over {\cal M}^7}X^2\Box\varphi , \label{L}\ee
where the sign before ${\varphi^5\over {\cal M}^5}\,X$ is
ghostlike, however, as will be showed that there are not the ghost
and gradient instabilities, since $Q>0$ and $c_s^2>0$. The
evolution of background is determined by \ba & &-{\varphi^5\over
{\cal M}^5}{\ddot \varphi}+3\left({7\over {\cal
M}^{10}}X^{5/2}+{8\over {\cal M}^7}H{\dot
\varphi}X\right){\ddot \varphi}\nonumber\\
&+& 3\left(-{\varphi^5\over {\cal
M}^5}+{7\over 2{\cal M}^{10}}X^{5/2}\right)H{\dot \varphi}\nonumber\\
&+& \left(-{5\varphi^4\over {\cal M}^5}+ {6{\dot H}{\dot
\varphi}^2\over {\cal M}^7}+{18{ H}^2{\dot \varphi}^2\over {\cal
M}^7}\right)X=0, \label{phi}\ea \be 3H^2 M_P^2 =\,
-\,{\varphi^5\over {\cal M}^5}\,X+{6\over {\cal M}^{10}}X^{7/2} +
{6\over {\cal M}^7}X {\dot \varphi}^3H
.\label{H}\ee  We require
${\varphi^5\over {\cal M}^5}X\simeq {6X^{7/2}\over {\cal M}^{10}}$
in Eq.(\ref{H}),
which gives \be \varphi=\varphi_* e^{-{\sqrt{2}\over 6^{1/5}}{\cal
M} (t_*-t)}. \label{ephi}\ee Thus $ H \simeq  {{\dot
\varphi}^5\over {\cal M}^7 M_P^2}$,
\ba a & \sim & e^{\int Hdt} = e^{{2 \varphi_*^5 {6}^{1/5}\over 15
{\cal M}^3
M_P^2}e^{-5{\sqrt{2}\over 6^{1/5}}{\cal M}(t_*-t)}} \label{amodel}\ea 
is induced. Thus for $\beta={5\sqrt{2}\over 6^{1/5}}{\cal M}$,
Eq.(\ref{model2}) is obtained, which is just required evolution.

We have $H{\dot \varphi}\ll {\ddot \varphi}$ for $\beta (t_*-t)\gg
1$, since \be H\sim {\dot \varphi}^5\sim e^{-5{\sqrt{2}\over
6^{1/5}}{\cal M} (t_*-t)}\ll e^{-{\sqrt{2}\over 6^{1/5}}{\cal M}
(t_*-t)}. \label{HH}\ee  Thus Eq.(\ref{phi}) is approximately \be
\left(-{\varphi^5\over {\cal M}^5}+{21\over {\cal
M}^{10}}X^{5/2}\right){\ddot \varphi}-{5\varphi^4\over {\cal
M}^5}X \simeq 0, \label{phiapp}\ee which is consistent with the
solution (\ref{ephi}).

The numerical solutions of Eqs.(\ref{phi}) and (\ref{H}) are
plotted in Fig.\ref{fig:aH}, which is consistent with
(\ref{amodel}) and will be applied to the numerical calculations
for the power spectrum of the adiabatical perturbation plotted in
Fig.3.


\subsection{The curvature perturbation}

$\cal R$ satisfies Eq.(\ref{uk}). We follow the definitions and
calculations in Ref.\cite{KYY}, and will only list the calculating
results in term of the Lagrangian (\ref{L}) with the solutions
(\ref{amodel}) and (\ref{ephi}) of the background and field. The
calculating steps are the same with section III.B in
Ref.\cite{Liu:2011ns}, which will be neglected here.

We have, for $\beta (t_*-t)\gg 1$,
\ba Q
&\simeq &{4 {\cal M}^3 M_P^2 6^{1/5}\over
\sqrt{2}\varphi_*^5}e^{{5\sqrt{2}\over 6^{1/5}}{\cal
M}(t_*-t)}\nonumber\\ &\sim & e^{\beta (t_*-t)}. \label{Q}\ea Thus
$Q\sim |\epsilon|
>0$, which is just required here, satisfies Eq.(\ref{e2}).
The $c_s^2$ is given by \ba c_s^2 & \simeq &
{16\over 6^{4/5}(14\sqrt{2}-3)}{{\cal M}^2\over \varphi_*^2
}e^{{2\sqrt{2}\over 6^{1/5}}{\cal M}(t_*-t)}\nonumber\\ & \sim &
e^{{2\over 5}\beta (t_*-t)}. \label{cs}\ea Thus $c_s^2>0 $ is also
just required, and satisfies Eq.(\ref{cs2}).
Thus there are not the ghost and gradient instabilities, the
effective field theory is healthy.

Thus the spectrum of $\cal R$ is scale invariant. The amplitude of
spectrum is given by Eq.(\ref{P1}) \be {\cal P}_{\cal R}^{1/2}\sim
{{\cal M}\over M_P}. \ee Thus ${\cal P}_{\cal R}^{1/2}\sim
10^{-5}$ requires ${\cal M}\sim 10^{-5} M_P$, which implies that
the only adjusted parameter in this model is set by the
observation. There is not other finetuning.

The universe will reheat after the ``emergence" or the slowly
expanding phase ends. The reheating mechanism has been discussed
in Refs.\cite{Liu:2011ns},\cite{PZhang}. Hereafter, the evolution
of hot ``big bang" model begins, the perturbation mode outside of
horizon will preserve constant.

\begin{figure}[t]
\includegraphics[width=7cm]{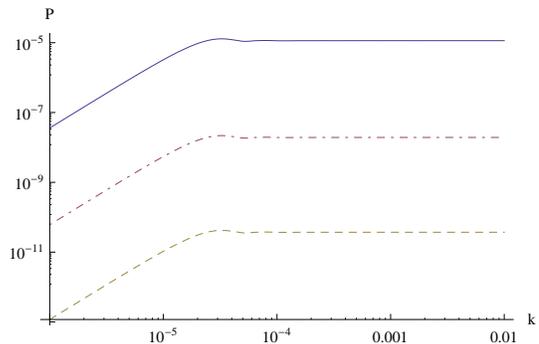}
\caption{The spectrum of curvature perturbation at different times
with respect to $k$, ${\cal M}=0.01$. The long dashed, dot dashed
and solid lines are the spectra at different times
$t\,=\,-100,-10,-1$, respectively. We see that the perturbation
spectrum is scale invariant, while the amplitude of perturbation
is increasing with the time, however, only is its amplitude
increasing but the shape of the spectrum is not altered.
}\label{fig:p}
\end{figure}

We can reformulate Eq.(\ref{uk}) as \be
u^{\prime\prime}_k+\left(c_s^2k^2-{ {{z}}^{\prime\prime}\over
{{z}}}\right)u_k=0, \label{uet}\ee where $'$ denotes the
differential with conformal time, and $u_k \equiv z{\cal R}_k$ and
$z\equiv a\sqrt{2M_P^2 Q}/c_s$. We numerically solve
Eq.(\ref{uet}) with the numerical solutions of Eqs.(\ref{phi}) and
(\ref{H}), and plot the evolution of the amplitude of the
adiabatical perturbation in Fig.\ref{fig:p}.

We can see that the perturbation spectrum is scale invariant,
which is consistent with analytical result. The amplitude of
perturbation is increasing with the time, however, only is its
amplitudes increasing but the shape of the spectrum is not altered
\cite{Piao0901},\cite{ZLP}.

\section{Discussion}

The idea of the emergent universe, i.e.the universe might
originate from a static state in the infinite past, is
interesting. It is generally thought that after the universe
emerges from the static state, a period of inflation will be
required, which sets the initial conditions of hot ``big bang"
evolution.

However, when the universe begins to deviate from static state, it
might slowly expanding. We parameterize the evolutions of this
slowly expanding phase or the ``emergence" into different classes,
and show that the scale invariant adiabatical perturbation could
be generated, i.e.the initial conditions of ``big bang" model
could be set, during the corresponding slowly expanding period.
After the slowly expanding phase ends, the universe reheats and
the evolution of hot ``big bang" model begins.


In the effective field theory setting, the physics yielding the
``emergent" phase has been not still understood well, since
generally the evolution of ``emergence" or the slow expansion
emerging from static state in infinite past requires $\epsilon\ll
-1$, as has been argued, and usually $\epsilon<0$ implies the
corresponding effective field theory has the ghost instability.
Here and in Ref.\cite{Liu:2011ns}, we have showed that
the evolutions of the ``emergence", which sets the initial
conditions of ``big bang" model, can be realized by applying
generalized Galileon Lagrangians, in which there is not the ghost
instability, the effective field theory is healthy.


In Refs.\cite{KYY},\cite{Seery}, the inflation scenario has been
implemented by applying the generalized Galileon, in which the
scale invariant spectrum of curvature perturbation can also
adiabatically induced.
After the universe emerges from the initial static state, if a
period of inflation happens, the initial conditions of hot ``big
bang" model will be set by the inflation scenario, as in original
Refs.\cite{Ellis:2002we},\cite{Ellis:2003qz}. This issue is also
interesting for exploring.




Here, we only bring one of all possible implements of the slow
expansion. In principle, there might be other effective actions
which could give similar results. It is obviously significant to
find a Lagrangian with a better theoretical motivation, which is
interesting for investigating.


\textbf{Acknowledgments} This work is supported in part by NSFC
under Grant No:10775180, 11075205, in part by the Scientific
Research Fund of GUCAS(NO:055101BM03), in part by National Basic
Research Program of China, No:2010CB832804.

\end{document}